\documentclass[12pt,twoside]{article}   
\usepackage[super,sort,comma]{natbib}

\usepackage{fancyhdr}		

\usepackage[section]{placeins}   %

\usepackage{graphicx}

\usepackage{amsmath}

\usepackage{soul} 
\usepackage{comment}
\usepackage{changepage}

\makeatletter \renewcommand\@biblabel[1]{$^{#1}$} \makeatother
\newcommand{\cen}[1]{\begin{center} #1 \end{center}}

\newcommand{\pz}{\phantom{0}}

\usepackage[mathlines]{lineno}

\usepackage{hyperref}
\hypersetup{ colorlinks,
    citecolor=blue,
    filecolor=blue,
    linkcolor=blue,
    urlcolor=blue
}

\usepackage{xcolor}

\definecolor{gray}{rgb}{0.6,0.6,0.6}
\definecolor{red}{rgb}{0.85,0,0}
\definecolor{green}{rgb}{0,0.85,0}
\definecolor{blue}{rgb}{0,0,0.85}
\definecolor{beige}{rgb}{0.92,0.87,0.78}
\definecolor{lightgreen}{rgb}{0.5,0.99,0.5}
\definecolor{lightcyan}{rgb}{0.5,0.99,0.99}
\definecolor{lightred}{rgb}{0.99,0.8,0.8}
\newcommand{\hg}[1]{{\sethlcolor{lightgreen}\hl{#1}}}

\sethlcolor{white}

\usepackage[all]{hypcap}    

\begin{document}


\cen{\sf {\Large {\bfseries A stopping criterion for iterative proton CT image reconstruction based on correlated noise properties } \\
\vspace*{10mm}
Ethan A. DeJongh$^1$, 
Alexander A. Pryanichnikov$^{2,3,\dagger}$,
Don F. DeJongh$^{1+}$,
Reinhard W. Schulte$^4$
}
\vspace{5mm}\\
$^1$ProtonVDA LLC, Naperville, IL 60563, USA\\
$^2$Moscow State University, Moscow, Russian Federation\\
$^3$Physical-Technical Center of Lebedev Physical Institute, Protvino, Russian Federation\\
$^4$Loma Linda University, Loma Linda, CA 92350, USA\\
\vspace{5mm}
}

\pagenumbering{roman}
\setcounter{page}{1}
\pagestyle{plain}
\cen{$^+$Author to whom correspondence should be addressed. email: fritz.dejongh@protonvda.com}
\cen{$^\dagger$ Now at Division of Biomedical Physics in Radiation Oncology, German Cancer Research Center (DKFZ), Im Neuenheimer Feld 280, Heidelberg, Germany}

\newpage



\pagenumbering{roman}
\setcounter{page}{2}

\begin{abstract}
\noindent {\bf Background:} Whereas filtered back projection algorithms for voxel-based CT image reconstruction have noise properties defined by the filter, iterative algorithms must stop at some point in their convergence and do not necessarily produce consistent noise properties for images with different degrees of heterogeneity. \\
{\bf Purpose:} A least-squares iterative algorithm for proton CT (pCT) image reconstruction converges toward a unique solution for relative stopping power (RSP) that optimally fits the protons. We present a stopping criterion that delivers solutions with the property that correlations of RSP noise between voxels are relatively low. This provides a method to produce pCT images with consistent noise properties useful for proton therapy treatment planning, which relies on summing RSP along {lines} of voxels. Consistent noise properties will also be useful for future studies of image quality using metrics such as contrast to noise ratio, and to compare RSP noise and dose of pCT with other modalities such as dual-energy CT. \\
{\bf Methods:} With simulated and real images with varying heterogeneity from a prototype clinical proton imaging system, we calculate average RSP correlations between voxel pairs in uniform regions-of-interest versus distance between voxels. We define a parameter
\textit{r}, the remaining distance to the unique solution relative to estimated RSP noise, and our stopping criterion is based on \textit{r} falling below a chosen value. \\
{\bf Results:} We find large correlations between voxels for larger values of
\textit{r}, and anticorrelations for smaller values. For
\textit{r} in the range of 0.5 to 1, voxels are relatively uncorrelated, and compared to smaller values of \textit{r} have lower noise with only slight loss of spatial resolution.  \\
{\bf Conclusions:} {Iterative algorithms not using a specific metric or rationale for stopping iterations may produce images with an unknown and arbitrary level of convergence or smoothing. We resolve this issue  
by stopping iterations of a least-squares iterative algorithm when} \textit{r} {reaches the range of 0.5 to 1.} This defines a pCT image reconstruction method with consistent statistical properties optimal for clinical use, including for treatment planning with pCT images.
\end{abstract}


\newpage     



\setlength{\baselineskip}{0.1cm}      

\pagenumbering{arabic}
\setcounter{page}{1}
\pagestyle{fancy}

\section{Introduction}

Treatment planning for proton therapy requires a three-dimensional (3D) map of the patient in terms of relative stopping power (RSP), relative to water. Currently, treatment planning systems start with single-energy x-ray CT images (SECT) with voxel values measured in Hounsfield units (HU), and convert these to RSP maps using scanner-specific calibrations. There is no consistent one-to-one relationship between HU and RSP for different tissues and materials, leading to significant range uncertainties, typically quantified as 3.5\% $\pm$ 1 mm for treatment planning purposes at most proton treatment centers \cite{Paganetti_2012}.  Experimental comparisons of RSPs in several biological tissues using SECT, dual-energy CT (DECT), proton CT (pCT) and helium CT (HeCT) show that DECT, pCT and HeCT have potential to improve RSP accuracy~\cite{baer-comparison}.  Comparison of SECT and pCT in fresh postmortem porcine structures~\cite{porcine} shows the largest discrepancies in compact bone and cavitated regions, and demonstrates the potential for pCT to be used for low-dose treatment planning with reduced margins.

CT image reconstruction typically utilizes filtered back-projection (FBP) algorithms although advancements in computer hardware as well as algorithms have led to the use of iterative algorithms as well~\cite{Stiller}. The ramp filter achieves the ideal FBP reconstruction with sharp features, but also high levels of noise in terms of fluctuations in voxel HU values.  Standard filters roll off at higher frequencies, achieving smoother images, at the cost of increased covariance between pixel values. The detectability of lesions is therefore not necessarily directly correlated with image variance~\cite{Lesion-det}, and it is important to consider noise correlations between voxels as well as noise variance when evaluating noise properties. Different filters  result in noise fluctuations on different length scales, which can be evaluated with the noise power spectrum (NPS). The integral of the NPS yields the noise variance of the CT image (variance relative to the true RSP, rather than to the local average). The International Commission on Radiation Units and Measurements (ICRU) recommends quantitative measurement of the NPS for meaningful comparisons between different scanners~\cite{ICRU}.

\begin{figure}
  \begin{center}
  \includegraphics[width=0.3\textwidth]{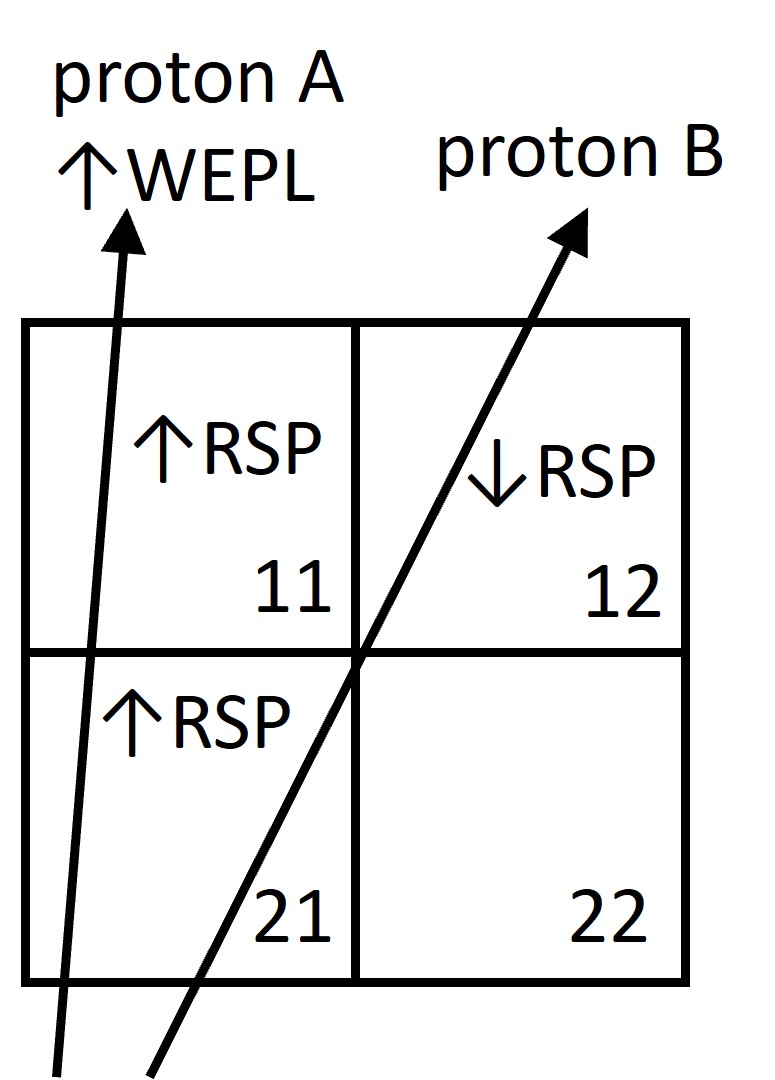}\\
  \caption{Proton WEPL noise can induce both correlations and anticorrelations in RSP noise.  In this example, {the WEPL of proton A helps determine the RSPs of voxels 21 and 11, and the WEPL of proton B helps determine the RSPs of voxels 21 and 12.} A fluctuation in the WEPL of proton A induces a correlated fluctuation in the RSPs of voxels 11 and 21. A fluctuation in voxel 21 combined with the WEPL constraint from proton B induces an anticorrelation with voxel 12. }\label{correlations}
  \end{center}
\end{figure}

Reconstructions of pCT images from list-mode pCT scanners use measurements of the trajectory and water equivalent path length (WEPL) of individual protons. 
The WEPL calibration of pCT range detectors can be traced to the range calibration of the proton therapy facility and the known water equivalent thickness (WET) of standard material blocks~\cite{bashkirov-range,dejongh2020technical}. A pCT image providing a good fit to a set of proton trajectories and WEPL measurements therefore constitutes a direct measurement of the 3D RSP map.
The WEPL measurement of a proton contributes to the RSP determination of the voxels touched by the trajectory of that proton, and in turn the RSP value of one voxel affects the fit of the image to the WEPLs of all the protons that touch that voxel. As illustrated in Fig.~\ref{correlations}, this interdependence can induce both correlations and anticorrelations in the RSP noise between voxels. 

A pCT image reconstruction algorithm must account for the curved proton trajectories resulting from multiple Coulomb scattering. A distance-driven FBP algorithm accounts for the curved trajectories by binning protons at each depth for the projections at each angle~\cite{Rit}. The convolution with the ramp filter for each projection uses binned data from the optimal depth at each angle for each voxel.  

Iterative algorithms fully account for the curved trajectories by adjusting the RSPs of the voxels touched by the protons to match the WEPLs of the protons. For their pCT and HeCT image reconstructions, B\"{a}r et al.~\cite{baer-comparison} used the DROP-TVS algorithm~\cite{drop-tvs}, which combines the block-iterative diagonally relaxed orthogonal projections (DROP) algorithm~\cite{Penfold} with a total variation superiorization (TVS) scheme that seeks feasible solutions with reduced image variance.  They applied the algorithm with a fixed relaxation parameter (i.e., step size) for 25 iterations, while noting that a higher number of iterations increased mean RSP accuracy, but also increased the image noise.

Least-squares iterative algorithms provide a unique solution that enables definitions of criteria to determine when to stop iterating~\cite{dejongh2020iterative,gmres-methods,gmres-stopping}. The least-squares method has additional features including the definition of an iteration vector that takes into account proton measurement uncertainties, the definition of an optimal step size for each iteration individually, the ability to simultaneously optimize the step sizes of many iterations, and the ability to divide the proton data into arbitrary numbers of blocks for parallel processing and use of graphical processing units. Similarly to the solutions with ramp filters in FBP methods, the least squares solutions have relatively sharp features and high noise. An iterative algorithm that starts with a smooth initial guess may, after first converging towards a smooth solution, produce noisier images as the number of iterations increases and the RSP values accommodate to the fluctuations in the WEPL measurements. The increase in noise with the number of iterations is known as ``semi-convergence''~\cite{gmres-methods} and leads to stopping criteria that balance noise with remaining distance to the unique solution.
The increase in noise with iteration number described in~Ref.{\setcitestyle{numbers}[\cite{baer-comparison}]} is typical of semi-convergence.

We have observed that the stopping criterion affects the level of correlations and anticorrelations in the RSP noise between voxels.
{We suggest that consistent and predictable noise properties of pCT images in terms of autocorrelation are advantageous for proton treatment planning. First, images with minimum autocorrelation will have the least impact on the precision and accuracy of the voxel sum of RSP values (WEPL), which are needed for proton range calculations, and, second, clinicians and treatment planners often request certain image characteristics such as high spatial resolution or low image noise for certain treatment planning tasks related to recognizing and segmenting tumor volumes and organs at risk. Consistently providing them  with these image properties, as determined by the stopping criterion, will thus be helpful.}

We present herein a least-squares stopping criterion with the property that correlations and anticorrelations between RSPs of voxels are relatively low, providing a method to produce pCT images with consistent noise properties useful for treatment planning.
We evaluate, for a set of example images with a range of heterogeneity in RSP, the level of correlations with the autocorrelation function, evaluated from the average correlation between voxel pairs in a uniform region-of-interest versus the distance between voxels.  The autocorrelation function and the NPS are a Fourier transform pair related by the Wiener-Khinchin theorem~\cite{Faulkner1984}. The resulting method will be useful for defining standard image reconstructions for NPS studies comparing the noise levels and imaging dose of pCT imaging with other modalities such as DECT.

\section{Materials and Methods}

\subsection{Images from the ProtonVDA prototype clinical pCT scanner}

We acquired pCT images with the prototype clinical ProtonVDA proton imaging system~\cite{dejongh2020technical}, positioned in a horizontal beam line equipped with PBS at the Northwestern Medicine Chicago Proton Center (NMCPC).  The proton imaging system acquires proton data in list mode, with tracking detectors to measure individual proton trajectories and a range detector to measure individual proton WEPLs. Each image acquisition uses three or four incoming proton energies, and 90 angles separated by $4^\circ$. For this project, we used reconstructed images~\cite{porcine} of a custom cylindrical phantom which incorporates a set of eight 4 cm high, 1.8 cm diameter tissue-equivalent cylindrical inserts, and of a sample of porcine pectoral girdle (shoulder) and ribs, with the ribs partially surrounding the shoulder, and the assembly of porcine tissues inserted into a blue wax cylinder with an inner radius of 20 cm. Ref.{\setcitestyle{numbers}[\cite{porcine}]} contains further details of the phantoms, proton data acquisition, and image reconstruction.

\subsection{Simulated images}

{The use of simulated images enabled us to determine the optimal scanning parameters (number of particles and angles) in terms of the noise metric, and to test the characteristics of correlations and anticorrelations before exploring them with the experimental datasets.}

We produced simulated images with the TOPAS software package\cite{TOPAS,TOPAS2}, an extension of GEANT4 for proton therapy tasks which incorporates a complete list of physical processes for proton interactions in the intermediate energy range. Electromagnetic processes include multiple Coulomb scattering, ionization by charged particles, and production of delta-electrons. Hadronic processes include elastic scattering of hadrons on the nuclei of the medium, and inelastic scattering of hadrons on nuclei with the production of secondary particles. {We used the physics lists FTFP\_BERT, G4HadronElasticPhysics and G4HadronInelasticQBBC.}

The detector model incorporates material for the tracking detectors, {including the segmentation of our tracking planes into stacks of 1 mm scintillating fibers}\cite{dejongh2020technical}.  {We placed the detector planes 30 cm apart, centered around the object being imaged.  The 1 mm pitch of the scintillating fibers determines the position resolution of the tracking detectors.  Our scintillating fiber stacks consist of two layers offset by 0.5 mm, allowing protons to be localized to 0.5 mm in each direction in each plane.}

{We model the WEPL resolution using our simulation of the average and the spread of the detector response for a grid of positions and residual ranges.}
We first generate simulated range detector output based on the position and residual range for each proton reaching the range detector.   {We then reconstruct the simulated WEPL from the simulated detector output and our standard residual range reconstruction algorithm}~\cite{dejongh2020technical}.

The proton beam model reproduces the NMCPC geometry including the positions and angles of the pencil beams at isocenter.  {The pencil beam scanning system has a focal length of 193.5 cm (horizontal) and 230.3 cm (vertical) from the scanning magnets to the isocenter plane.}

We produced a variety of simulated pCT images of a simple cylindrical water phantom placed between the detector tracker planes.
The geometry of the simulated water phantom matched that of the custom cylindrical phantom, with a height of 4 cm and diameter of 18 cm. We used three proton energies in the simulation, 118, 160, and 195 MeV, to be similar to real data from our prototype system. Because the uniform cylinder has the same features from all sides, we were able to generate all of the simulated protons from one direction and then assign them randomly to a chosen number of different rotation angles. This allowed us to alter the number of angles and number of protons per angle for different pCT images without repeating the time-intensive simulation.

\subsection{Summary of the least-squares iterative method}

As described in detail in~Ref.{\setcitestyle{numbers}[\cite{dejongh2020iterative}]}, we reconstruct pCT images using a least-squares iterative method with a stopping criterion based on RSP noise.  We briefly describe here the main features of the method, which solves the matrix equation $Ax = b$ for $x$, where $b$ is a vector with one entry per proton containing the WEPL measurements for each proton, $x$ is a vector with one entry per voxel containing the RSP for that voxel, and $A$ is a matrix with one row for each proton and one column for each voxel, where each entry contains the chord length of the proton trajectory through the voxel. The product $Ax$ is the vector of calculated WEPL values for each proton for a possible solution $x$.

Since there are many more protons than voxels, the proton deviation vector $d_p$, where
\begin{equation}
    d_p = Ax-b,
\end{equation}
will not be zero for any solution.  Instead, we define the voxel deviation vector $d_v$ as a weighted average of the $d_p$ of all the protons going through each voxel, with 
\begin{equation}
    d_v = \bar{A}^T d_p
\end{equation}
where $\bar{A}^T = V^{-1}A^T$, 
$V^{-1} = \textit{diag}\left({1}/{\sum_j \alpha^T_{ij}}\right)$, and $\alpha^T_{ij}$ are the elements of $A^T$.
We use $d_v$ to iterate towards an optimal solution for $x$:  
\begin{equation}
    x \rightarrow x - \lambda d_v, 
\end{equation}
where the relaxation parameter $\lambda$ defines the choice of step size. We are able to simultaneously optimize the step sizes for multiple iterations, with the optimum based on to minimizing either $d_p \cdot d_p$ or $d_v \cdot d_v$ after the iterations, and have found that a good strategy is to alternate between these choices. The use of $d_v$ as the iteration vector helps ensure a uniform convergence rate across the image, a useful property when choosing a stopping point.
For the unique least squares solution, $d_v = 0$ and $\bar{A}^TAx = \bar{A}^T b$.  For this solution, the average calculated WEPL of protons touching a voxel is equal to the average measured WEPL of those protons.

The $d_p$ vector can never reach zero, since entries for individual protons will fluctuate around zero according to uncertainties in the WEPL measurement, as well as from the effect of uncertainties in the proton trajectory on the calculated WEPL. For a solution vector $x$, the standard deviation of $d_p$ therefore provides an estimate of the average proton noise
{$\sigma_p$. With $N_{pv}$ as the average number of protons touching a voxel, as obtained from the data, we define the estimated average voxel precision as:}
\begin{equation}\label{sigmav}
    \sigma_v = \frac{\sigma_p}{\bar{\alpha}\sqrt{N_{pv}}}
\end{equation}
{where $\bar{\alpha}$ is the average chord length of a proton through a voxel.  For our purposes, we can simply approximate this as the length of the side of a voxel. We then compare $\sigma_v$ to} the root-mean-square (r.m.s.) of $d_v$, the remaining distance to the least-squares solution\cite{dejongh2020iterative}.  We stop iterating if this remaining distance is small compared with the estimated RSP precision. We define $r$ as the ratio
\begin{equation}\label{stopping}
 r = \frac{\text{r.m.s.}\ d_v}{\sigma_v}
\end{equation}
and our stopping criterion is based on $r$ falling below a chosen value. This stopping criterion does not depend on knowledge of the true RSP distribution, and generally applies to any object being imaged.

\subsection{Definitions of the autocorrelation functions}

While our stopping criterion applies to any image, analysis of the statistical properties of a particular image does require some knowledge of the true RSP distribution for that image. 
We perform our noise analyses within uniform regions of interest (ROI) with constant RSP across voxels.  In the case of the simulated cylindrical phantom, the entire cylinder consists of a uniform volume.  In the case of the custom cylindrical phantom, the blue wax is uniform and each insert individually is uniform.  In the case of the pork shoulder with ribs, we use the reconstructed image to select uniform regions with a single tissue type.

We label voxels with the index numbers $i, j, k$ along the $x, y, z$ axes, and calculate the average RSP, $\overline{RSP}$ in each ROI. 
In this coordinate system, the object being imaged rotates around the $y$ (vertical) axis, and the protons travel mostly in the $x-z$ plane with some divergence into the $y$ direction. Our ROIs consist of cubes with 50 voxels on a side. In the case of the custom cylindrical phantom, the ROI consists of a volume with $50 \times 30 \times 50$ voxels in the $x$, $y$ and $z$ directions.

The variance is the square of the standard deviation $\sigma$ within an ROI:
\begin{equation}
    \sigma^2 = \frac{1}{N-1} \sum_{i,j,k} (RSP_{ijk}-\overline{RSP})^2 .
\end{equation}

We calculate the autocorrelation functions $\rho$ as the average correlation coefficient of pairs of voxels with a separation $\delta$, along $x, y$ or $z$, within an ROI:

\begin{align}
    \rho_{x,\delta} &= \frac{1}{N} \sum_{i,j,k} \frac{(RSP_{ijk}-\overline{RSP})(RSP_{(i+\delta)jk}-\overline{RSP})}{\sigma^2}\\
    \rho_{y,\delta} &= \frac{1}{N} \sum_{i,j,k} \frac{(RSP_{ijk}-\overline{RSP})(RSP_{i(j+\delta)k}-\overline{RSP})}{\sigma^2}\\
    \rho_{z,\delta} &= \frac{1}{N} \sum_{i,j,k} \frac{(RSP_{ijk}-\overline{RSP})(RSP_{ij(k+\delta)}-\overline{RSP})}{\sigma^2}
\end{align}

\section{Results}

\begin{figure}
    \begin{center}
    \includegraphics[width=0.44\textwidth]{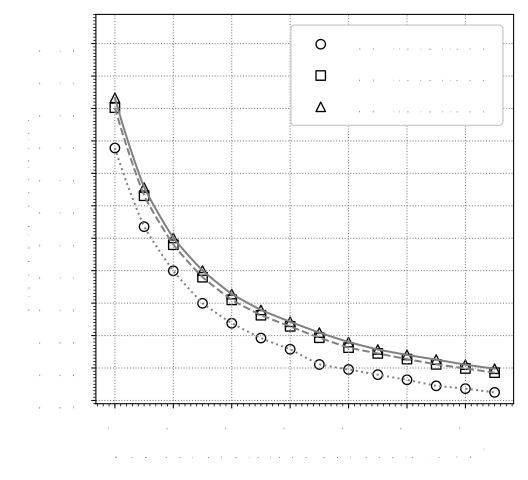}
    \caption{Standard deviation of the RSPs in an ROI in the center of the simulated uniform cylindrical phantom.  Each simulation used 90 angles separated by 4$^\circ$, and the number of protons per angle indicated on the horizontal axis. Each image reconstruction used the fixed number of iterations indicated in the legend. }\label{it_vs_protons}
    \end{center}
\end{figure}

\begin{figure}
    \begin{center}
    \includegraphics[width=0.44\textwidth]{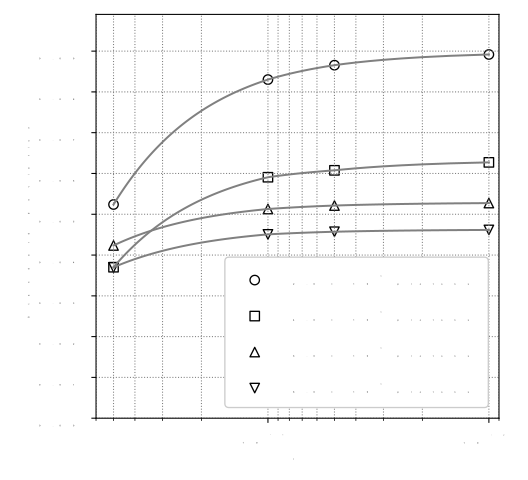}
    \caption{Standard deviation of the RSPs in an ROI in the center of the simulated uniform cylindrical phantom.  Each simulation used 90 angles separated by 4$^\circ$, and the number of protons per angle indicated in the legend. Each image reconstruction stopped after dropping below the $r$ value indicated on the horizontal axis.}\label{std_dev_vs_r}
    \end{center}
\end{figure}

Fig.~\ref{it_vs_protons} shows the standard deviation of RSPs in an ROI in the center of the simulated uniform cylindrical phantom for several simulations with different numbers of protons, and several image reconstructions with different fixed numbers of iterations. As expected, the standard deviation decreases as the square root of the number of events.  However, the image reconstructions show {behavior typical of semi-convergence}, with the standard deviation increasing with the number of iterations until reaching a maximum value as illustrated in Fig.~\ref{std_dev_vs_r}. The standard deviation increases as $r$ decreases, reaching a plateau as $r$ falls below 1 and the solution becomes closer to the unique least-squares solution.

\begin{figure}
    \begin{center}
    \includegraphics[width=0.44\textwidth]{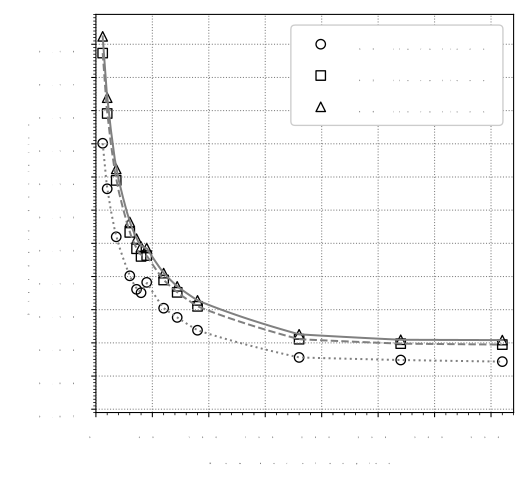}
    \caption{Dependence of the standard deviation of the RSP values on the number of projection angles.}\label{it_vs_angles}
    \end{center}
\end{figure}

While we have based our results on pCT image reconstructions using 90 angles separated by $4^\circ$ each, we have investigated how the noise of our images depends on the number of angles used, for a fixed total number of protons.  As shown in Fig.~\ref{it_vs_angles}, we have found that the use of 180 angles separated by $2^\circ$ is better with $\approx 10\%$ lower standard deviation, and we may in future studies use 180 angles. The noise increases rapidly with the use of fewer angles.

\begin{figure}
    \begin{center}
    \includegraphics[width=0.44\textwidth]{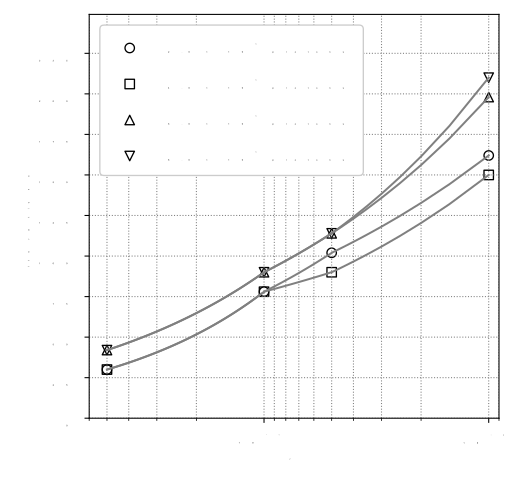}
    \caption{Number of iterations required to reach a given $r$ value for image reconstructions of the simulated uniform cylindrical phantom.  Each simulation used 90 angles separated by 4$^\circ$, and the number of protons per angle indicated in the legend.}\label{it_vs_r}
    \end{center}
\end{figure}

Fig.~\ref{it_vs_r} relates the value of $r$ achieved to the number of iterations applied for the example of the simulated uniform cylindrical phantom, using a variety of numbers of protons per angle.  
Since $r$ is based on the noise level (i.e. on the number of protons), for a given $r$ we need to iterate closer to the ground truth when using more protons, therefore we expect the number of iterations needed for a given $r$  to increase with the number of protons used.  
Encouragingly, and as shown in Fig. 5, our experience is that this dependence is not very strong.  However, iterating  towards closer proximity to the least-squares solution can carry a high computational cost, with the number of iterations needed approximately linear in -log($r$) {as seen in Fig. 5.  Once $r$ is significantly less than 1, there is relatively little change in the image relative to the noise as iterations continue, and in the following figures we use the solution at $r=0.2$ as an approximation to the least-squares solution.}

\begin{figure}
    \begin{center}
    \includegraphics[width=1.00\textwidth]{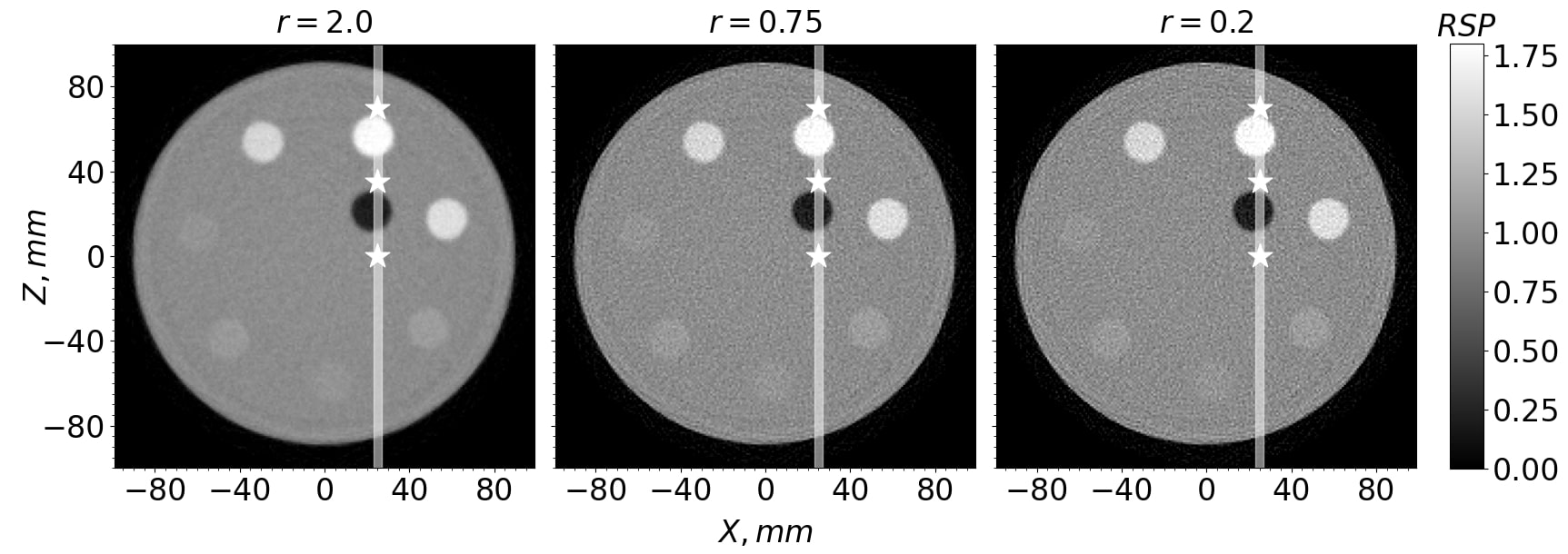}
    \caption{pCT slice, 1 mm thick, of the custom cylindrical phantom for three $r$ values. Star-shaped markers on the white line show the positions for the WET calculations listed in Table ~\ref{table:1}.}\label{phantom_slice_line}
    \end{center}
\end{figure}

\begin{figure}
    \begin{center}
    \includegraphics[width=1.00\textwidth]{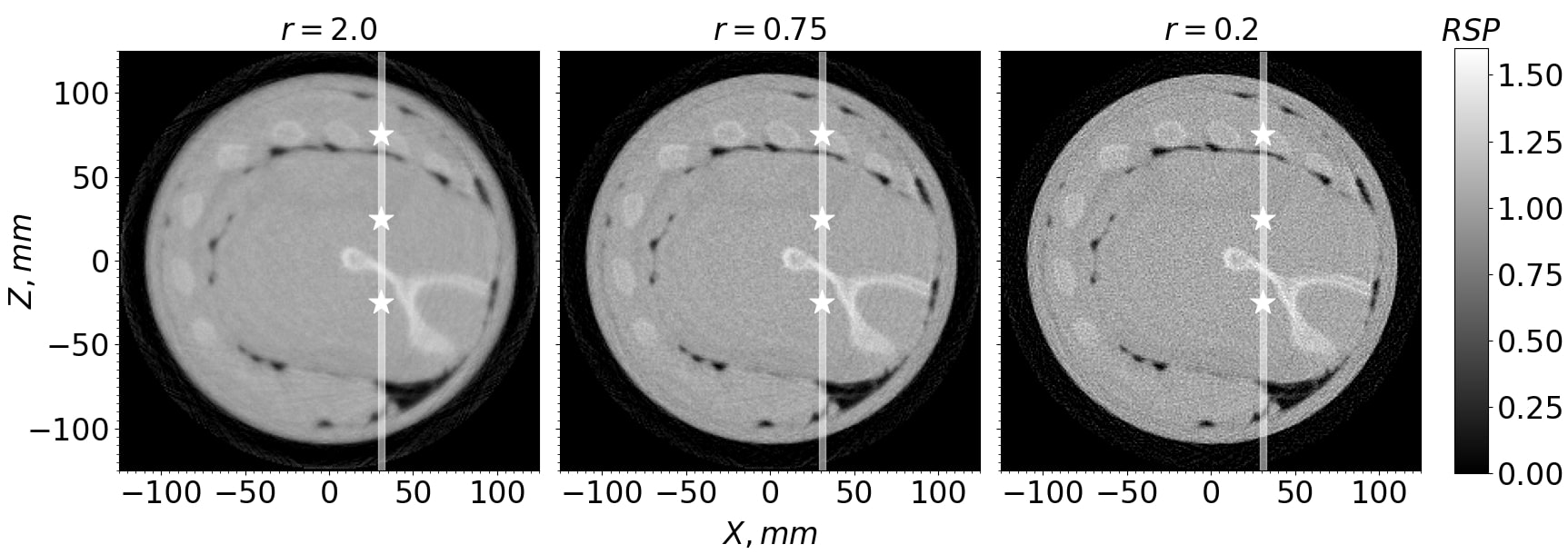}
    \caption{pCT slice, 1 mm thick, for the pork shoulder with ribs for three $r$ values. Star-shaped markers on the white line show the positions for the WET calculations listed in Table ~\ref{table:1}.}\label{pork_slice_line}
    \end{center}
\end{figure}

\begin{figure}
    \begin{center}
    \includegraphics[width=0.44\textwidth]{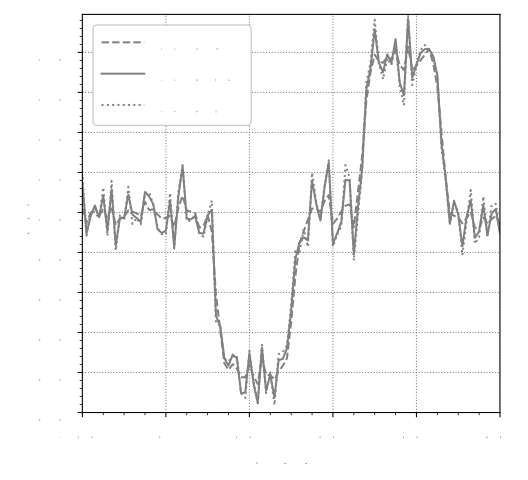}
    \caption{RSP profile along the white line in the slices shown in Figs.~\ref{phantom_slice_line}. }\label{phantom_line_r}
    \end{center}
\end{figure}

\begin{figure}
    \begin{center}
    \begin{minipage}[h]{1\linewidth}
    \center\includegraphics[width=0.44\textwidth]{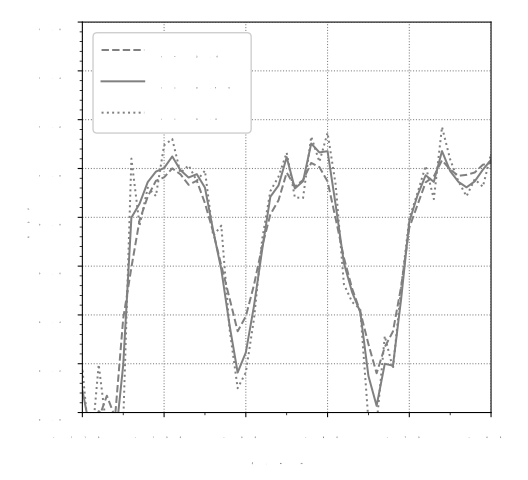}
    \end{minipage}
    \vfill
    \begin{minipage}[h]{1\linewidth}
    \center\includegraphics[width=0.44\textwidth]{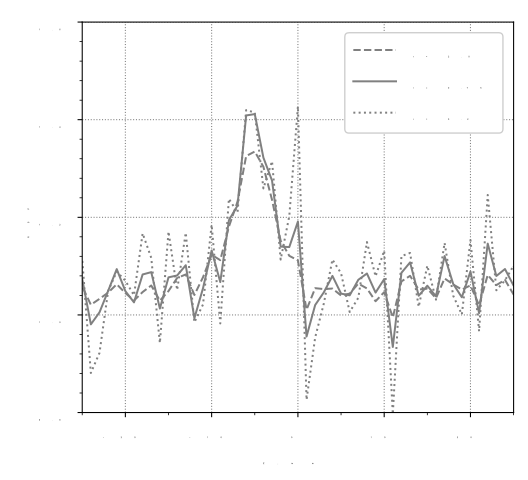}
    \end{minipage}
    \vfill
    \begin{minipage}[h]{1\linewidth}
    \center\includegraphics[width=0.44\textwidth]{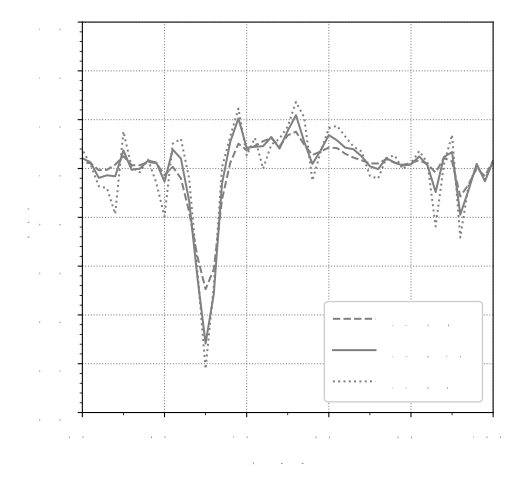}
    \end{minipage}
    \caption{RSP profiles along the white line in the slices shown in Figs.~\ref{pork_slice_line}.  }\label{pork_line_r}
    \end{center}
\end{figure}

Fig.~\ref{phantom_slice_line} shows an example of a pCT slice of the custom cylindrical phantom for three different $r$ values, 2.0, 0.75 and 0.2. Fig.~\ref{pork_slice_line} shows an example of a pCT slice of the pork ribs with shoulder for the same $r$ values. Fig.~\ref{phantom_line_r} shows a profile of RSP values along the lines through the slices in Fig.~\ref{phantom_slice_line}. Similarly, Fig.~\ref{pork_line_r} shows profiles of RSP values along the lines through the slices in Fig.~\ref{pork_slice_line}.
The images with $r=0.2$ are much closer to the least-squares solution and show a higher noise level {but also improved spatial resolution.  In contrast, the images for $r=2.0$ have less noise but the spatial resolution deteriorates.}
\hl{For a given value of $r$, the standard deviations of the RSP values within the inserts are very similar for each insert.}

We have found that WET values derived from the pCT images are usually not very sensitive to $r$, since the WET across a boundary with a sharp RSP differential does not depend strongly on the spatial resolution of the boundary. 
\hl{Thus, the exact choice of $r$ is not critical to the accuracy of proton range calculations.}
Table~\ref{table:1} shows some examples for the custom cylindrical phantom and the pork shoulder and ribs, and the WET values vary only by a few tenths of a mm with $r$. 
There may be exceptions for cases with trajectories grazing boundaries or starting or ending near boundaries, which will be investigated in future work. 

\begin{table}
\caption{Total WET along the white lines {in the +Z direction} in Fig.~\ref{phantom_slice_line} and  Fig.~\ref{pork_slice_line}, starting from the bottom and ending at the positions indicated by the star-shaped markers.}\label{table:1} 
\begin{center}
\begin{tabular}{ |c|c|c| } \hline
$r$ & WET (mm) & WET (mm)\\
 & Custom cylindrical phantom & Pork shoulder and ribs\\
& bottom\pz\pz middle\pz\pz top\pz\pz\pz & bottom\pz\pz middle\pz\pz top\pz\pz\pz \\ 
\hline
 2.0              & 86.57\pz\pz\pz\pz 121.16\pz\pz 155.58 & 82.61\pz\pz\pz\pz 136.81\pz\pz 184.88 \\ 
 1.0              & 86.79\pz\pz\pz\pz 121.29\pz\pz 155.84 & 82.52\pz\pz\pz\pz 137.02\pz\pz 184.69 \\
 0.75             & 86.81\pz\pz\pz\pz 121.25\pz\pz 155.78 & 82.42\pz\pz\pz\pz 137.04\pz\pz 184.64 \\
 0.5              & 86.88\pz\pz\pz\pz 121.25\pz\pz 155.83 & 82.22\pz\pz\pz\pz 136.95\pz\pz 184.50 \\
 0.2              & 86.95\pz\pz\pz\pz 121.33\pz\pz 155.95 & 82.01\pz\pz\pz\pz 136.83\pz\pz 184.47 \\
\hline
\end{tabular}
\end{center}
\end{table}

\begin{figure}
    \begin{center}
    \includegraphics[width=1.00\textwidth]{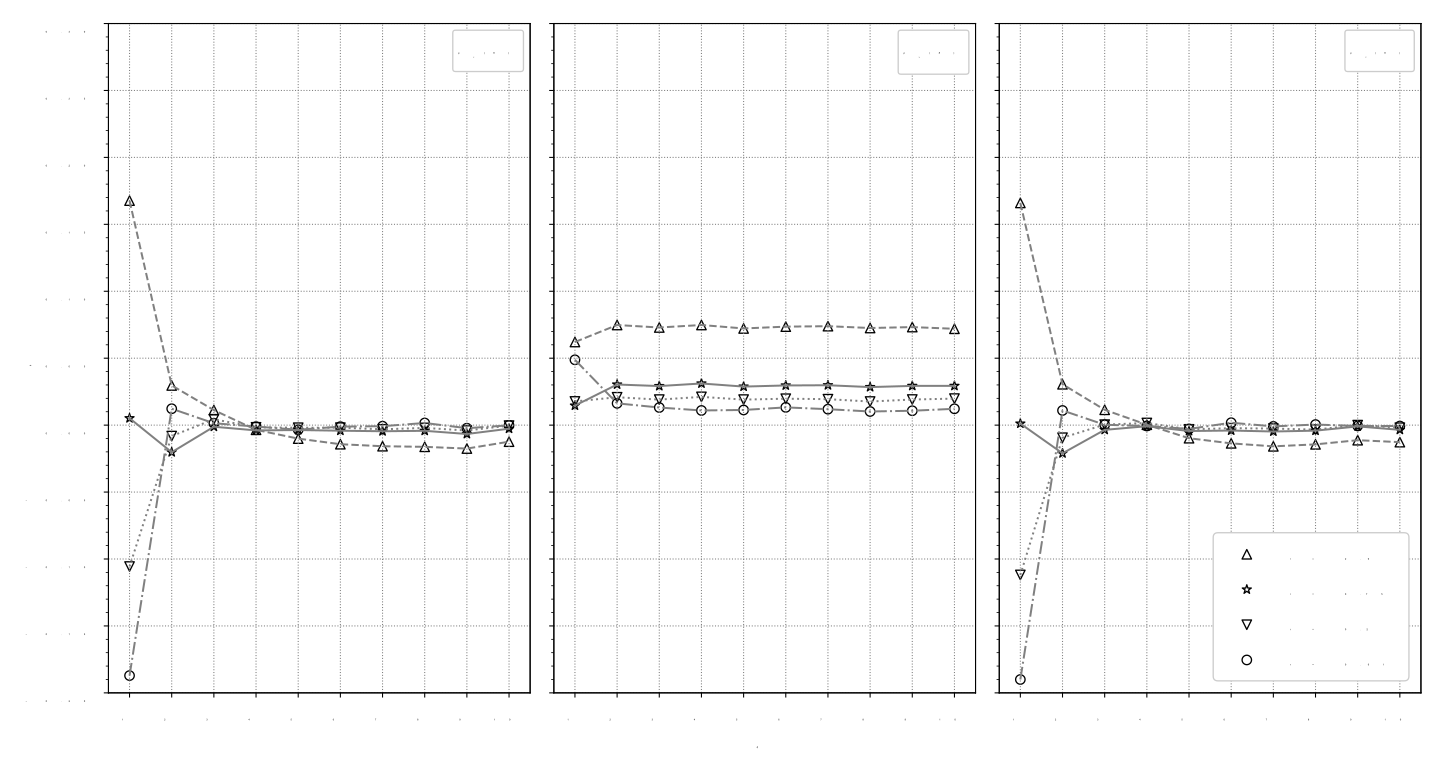}
    \caption{Autocorrelation functions $\rho_x$ (left) $\rho_y$ (middle) $\rho_z$ (right) versus the separation $\delta$ between voxels within a ROI of the simulated cylindrical water phantom for the $r$ values in the legend.}\label{simcyl_corr}
    \end{center}
\end{figure}

\begin{figure}
    \begin{center}
    \includegraphics[width=1.00\textwidth]{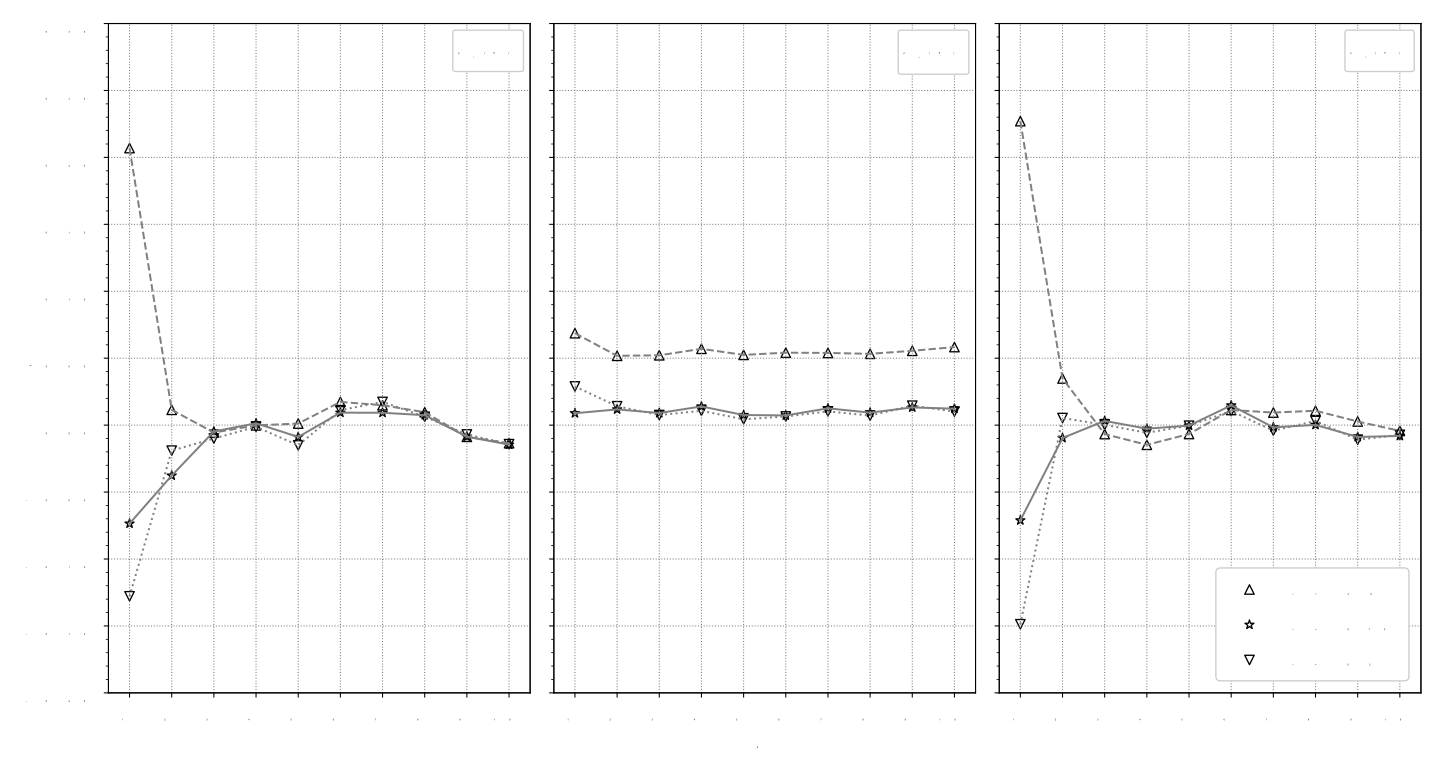}
    \caption{Autocorrelation functions $\rho_x$ (left) $\rho_y$ (middle) $\rho_z$ (right) versus the separation $\delta$ between voxels within a ROI of the custom cylindrical phantom for the $r$ values in the legend.}\label{phantom_corr}
    \end{center}
\end{figure}

\begin{figure}
    \begin{center}
    \includegraphics[width=1.00\textwidth]{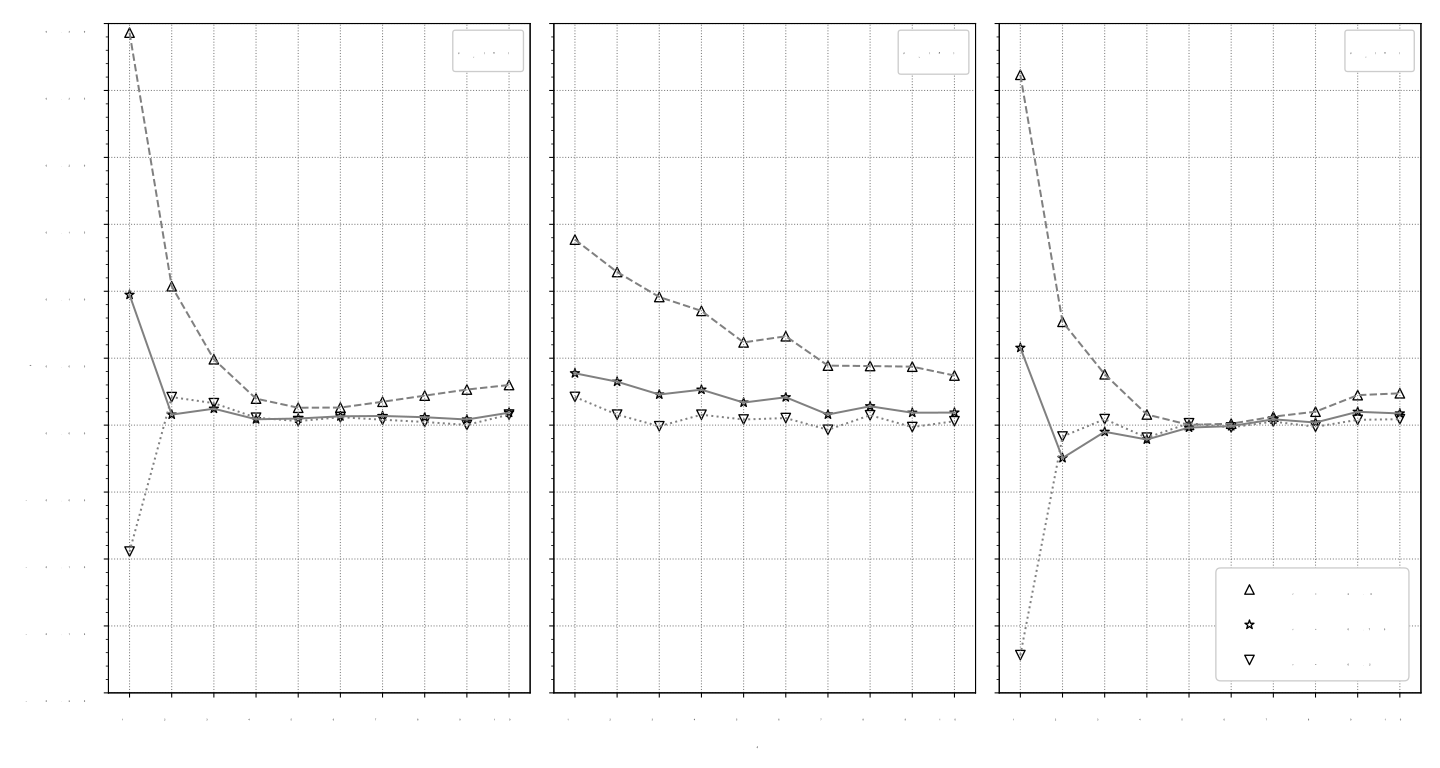}
    \caption{Autocorrelation functions $\rho_x$ (left) $\rho_y$ (middle) $\rho_z$ (right) versus the separation $\delta$ between voxels within a ROI of the pork shoulder and ribs for the $r$ values in the legend.}\label{pork_corr}
    \end{center}
\end{figure}

Fig.~\ref{simcyl_corr} shows the autocorrelation functions for the simulated cylindrical water phantom, Fig.~\ref{phantom_corr} for the custom, cylindrical phantom, and Fig.~\ref{pork_corr} for the pork shoulder and ribs.  Larger values of $r$ show positive correlations, while smaller values of $r$ show anticorrelations.  Although these images have much different heterogeneity, in each case stopping in the range $0.5 < r < 1$ produces minimal correlations while producing images almost as sharp as the least squares image.  In this range, the distance to the least squares solution is similar to the estimated statistical precision of the RSP values.

\section{Discussion}

In iterative reconstruction of pCT images, each iteration carries a high computational cost. A stopping criterion is essential for limiting this cost but must also consistently produce reasonable image quality in terms of noise and spatial resolution. Previous approaches typically stopped after a fixed number of iterations and often applied smoothing techniques such as TVS. Our approach uses a stopping criterion based on objective metrics of image quality.  {Although this consistency is not mathematically guaranteed, we expect our method to be generally applicable to any object being imaged, and have found good results} with a variety of real and simulated images with a range of heterogeneity.

Several stopping criteria have been previously proposed based on the $d_p$ vector~\cite{gmres-stopping}, exploiting the property that $d_p \cdot d_p$ reaches a mininum for the least squares solution. However, the value of this minimum may not be well known, and these criteria often depend sensitively on an understanding of the WEPL resolution for the protons,  as well as on uncertainties on the calculated WEPL from uncertainties in the proton trajectory, especially for more heterogeneous objects. Our stopping criterion is based on the size of the {the r.m.s. of} $d_v$ relative to image noise~\cite{dejongh2020iterative}, exploiting the property that for the least squares solution $d_v \cdot d_v = 0$, the value of the minimum is thus known, and an estimate of the expected image noise is sufficient for a robust comparison to the size of {the r.m.s. of} $d_v$.

The choice of filter in FBP reconstruction of CT images has the benefit of producing images with predictable statistical properties in terms of NPS, with the standard filter reducing high-frequency noise relative to the ramp filter. Our pCT image reconstruction exhibits the phenomenon of semi-convergence commonly observed with the use of iterative algorithms, with noise increasing with additional iterations after first converging from the initial starting guess. In our analysis, semi-convergence is a consequence of converging towards a least-squares fit to noisy data when starting with a smooth initial starting guess, where the noise arises from the WEPL resolution for the protons.  

Similarly to the ramp filter in FBP, the least squares solution produces images with high-frequency noise.   Our stopping criterion has benefits similar to those of FBP in terms of producing images with consistent statistical properties.  {For the cases we have tested, which include a set of images with a wide range of heterogeneity, we have found} we can stop at a point where statistical correlations between voxels, (or equivalently, the autocorrelation function), are relatively low and spatial resolution is relatively sharp.  {For special cases requiring the best spatial resolution, it is possible to further iterate to lower values of $r$ such as $r=0.2$, in analogy to the bone filter in x-ray CT reconstruction.}

{The use of our stopping criterion has the benefit of producing images with noise levels that realistically reflect the precision of the data for treatment planning, instead of being an arbitrary consequence of the choice of parameters of the iterative algorithm with an unknown level of smoothing. Also, our stopping criterion provides a basis for quantitative comparions between imaging modalities.  For example, DECT images often appear noisy compared to pCT images}~\cite{baer-comparison}, but a fair comparison should take into account the NPS (the Fourier transform of the autocorrelation function), any smoothing, and the delivered dose for each modality. Future work will focus on such quantitative comparisons.


\section{Conclusions}

As proton imaging moves into clinical use and pCT imaging becomes available for treatment planning, robust image reconstruction with consistent statistical properties will be essential.  
{Iterative algorithms not using a specific metric or rationale for stopping iterations may produce images with an unknown and arbitrary level of convergence or smoothing. We resolve this issue  
by stopping iterations of a least-squares iterative algorithm when} \textit{r} {reaches the range of 0.5 to 1.} 
This defines an
 algorithm with a stopping criterion that produces pCT images with consistent noise properties.  The stopping criterion applies generally to any object being imaged without prior knowledge of RSP, and is robust without relying on a detailed understanding of WEPL resolution.
The resulting images have relatively low statistical correlations between voxels, while maintaining good spatial resolution.  Future work will use this method with analysis of NPS to compare image noise and dose of pCT with other modalities such as dual-energy CT.

\section*{Data Availability Statement}

The data that support the findings of this study are available from the corresponding author upon reasonable request.

\section*{Author Contributions} Don F. DeJongh and Ethan A. DeJongh devised the stopping criterion concept. Reinhard W. Schulte provided feedback on clinical requirements for image characteristics. Alexander A. Pryanichnikov designed and implemented the data analysis. Don F. DeJongh, Ethan A. DeJongh, and Reinhard W. Schulte planned and implemented the acquisition of experimental and simulated data.  All authors provided critical feedback, shaped the research, and wrote the final manuscript.

\section*{Acknowledgments}

Research reported in this publication was supported by the National Cancer Institute of the National Institutes of Health under award number R44CA243939. Alexander A. Pryanichnikov acknowledges the Ministry of Science and Higher Education of Russian Federation (project number 075-15-2021-1347) for providing research support in the development of a software package for automating the calculation of the statistical characteristics of proton imaging.

\section*{Conflict of Interest Statement}

The authors have intellectual property rights to the innovations described in this paper. Don F. DeJongh is a co-owner of ProtonVDA LLC.

\clearpage


\section*{References}

\vspace*{-20mm}





\bibliography{./pCT_noise}      



\bibliographystyle{./medphy.bst}    


\end{document}